\documentclass{aa}
\usepackage{txfonts}
\usepackage{graphicx}
\usepackage[latin1]{inputenc}
\begin{document}
\def\Hp{H\,{\sc{ii}}}
\def\SII{[S\,{\sc{ii}}]}
\def\FeII{[Fe\,{\sc{ii}}]}
\def\HI{H\,{\sc{i}}}
\def\fs{\hbox{$.\!\!^{\rm s}$}}
\def\fdg{\hbox{$.\!\!^\circ$}}
\def\farcm{\hbox{$.\mkern-4mu^\prime$}}
\def\farcs{\hbox{$.\!\!^{\prime\prime}$}}
\def\arcmin{\hbox{$^\prime$}}
\def\arcsec{\hbox{$^{\prime\prime}$}}
\def\sun{\hbox{$\odot$}}
\def\degr{\hbox{$^\circ$}}
\def\h{\hbox{$^{\reset@font\r@mn{h}}$}}
\def\m{\hbox{$^{\reset@font\r@mn{m}}$}}
\def\s{\hbox{$^{\reset@font\r@mn{s}}$}}

\def\msol{\hbox{\kern 0.20em $M_\odot$}}
\def\lsol{\hbox{\kern 0.20em $L_\odot$}}
\def\smu{\hbox{\kern 0.20em s$^{-1}$}}
\def\kms{\hbox{\kern 0.20em km\kern 0.20em s$^{-1}$}}
\def\cmmt{\hbox{\kern 0.20em cm$^{-3}$}}
\def\cmmd{\hbox{\kern 0.20em cm$^{-2}$}}
\def\K{\hbox{\kern 0.20em K}}
\def\pc{\hbox{\kern 0.20em pc}}
\def\pcpd{\hbox{\kern 0.20em pc$^{2}$}}
\def\pcmu{\hbox{\kern 0.20em pc$^{-1}$}}
\def\twco{\hbox{${}^{12}$CO}}
\def\twcotwo{\hbox{${}^{12}$CO(2-1)}}
\def\thco{\hbox{${}^{13}$CO}}
\def\thcotwo{\hbox{${}^{13}$CO(2-1)}}
\def\thcoone{\hbox{${}^{13}$CO(1-0)}}
\def\ceio{\hbox{C${}^{18}$O}}
\def\ceiotwo{\hbox{C${}^{18}$O(2-1)}}
\def\ceioone{\hbox{C${}^{18}$O(1-0)}}
\def\cs{\hbox{CS}}
\def\csthree{\hbox{CS(3-2)}}
\def\cstwo{\hbox{CS(2-1)}}
\def\csfive{\hbox{CS(5-4)}}
\def\cts{\hbox{C${}^{34}$S}}
\def\ctsthree{\hbox{C${}^{34}$S(3-2)}}
\def\ctstwo{\hbox{C${}^{34}$S(2-1)}}
\def\htwo{\hbox{H${}_2$}}
\def\h13cop{\hbox{H$^{13}$CO$^{+}$}}
\def\halpha{\hbox{H$\alpha$ }}
\def\hcop{\hbox{HCO$^{+}$}}
\newcommand{\jonetozero}{\hbox{$J=1\rightarrow 0$}}
\newcommand{\jtwotoone}{\hbox{$J=2\rightarrow 1$}}
\newcommand{\jthreetotwo}{\hbox{$J=3\rightarrow 2$}}
\newcommand{\jfourtothree}{\hbox{$J=4\rightarrow 3$}}
\newcommand{\jfivetofour}{\hbox{$J=5\rightarrow 4$}}
\title{Triggered massive-star formation on the borders of 
       Galactic H\,{\Large II} regions }
\subtitle{III. Star formation at the periphery of Sh2-219 \thanks{Based on observations 
	obtained at the  
       IRAM Observatory, Spain, at the Observatoire de Haute-Provence, France, at the 
       Telescopio Nazionale Galileo, Canary Islands, and at the VLA, USA.}}
\author{L.~Deharveng\inst{1}
          \and
        B.~Lefloch\inst{2}
          \and 
        F.~Massi\inst{3}
          \and
        J.~Brand\inst{4}
          \and
        S.~Kurtz\inst{5}
           \and
        A.~Zavagno\inst{1}
           \and
        J.~Caplan\inst{1}
        }
 
 \offprints{L.~Deharveng}

\institute{
      Laboratoire d'Astrophysique de Marseille, 2 place Le Verrier, 
      13248 Marseille Cedex 4, France, lise.deharveng@oamp.fr
       \and
       Laboratoire d'Astrophysique de l'Observatoire de Grenoble, 414 rue de
       la Piscine, BP 53, 38041 Grenoble Cedex 9, France
       \and
       INAF-Osservatorio Astrofisico di Arcetri, Largo E.~Fermi 5, 50125 Firenze, Italy
       \and
       INAF-Istituto di Radioastronomia, Via Gobetti 101, 40129 Bologna, Italy
       \and
       Centro de Radioastronom\'{i}a y Astrof\'{i}sica, UNAM, Apartado Postal 3-72, 58089, Morelia, 
       Michoac\'an, M\'exico
        }

\date{Received; accepted }

\abstract{Massive-star formation triggered by the expansion of \Hp\ regions.}
{To understand if sequential star formation is taking place at the periphery of 
the \Hp\ region Sh2-219.} 
{We present \twco~\jtwotoone\ line  observations of this region, obtained at the IRAM 30-m 
telescope (Pico Veleta, Spain).}
{In the optical, Sh2-219 is spherically symmetric 
around its exciting star; furthermore it is surrounded along three quarters of 
its periphery by a ring of atomic hydrogen. This spherical symmetry breaks down  
at infrared and millimetre wavelengths. A molecular cloud of about 2000\msol\  
lies at the southwestern border of Sh2-219, in the \HI\ gap. Two molecular 
condensations, elongated along the ionization front, probably result from the 
interaction between the expanding \Hp\ region and the molecular cloud. In this 
region of interaction there lies a cluster containing many highly reddened stars, 
as well as a massive star exciting an ultracompact \Hp\ region. 
More surprisingly, the brightest parts of the molecular cloud form a 
`chimney', perpendicular to the ionization front. This chimney is closed 
at its south-west extremity by H$\alpha$~walls, thus forming a cavity. 
The whole structure is 7.5~pc long. A luminous H$\alpha$ emission-line 
star, lying at one end of the chimney near the ionization 
front, may be responsible for this structure. Confrontation of the observations 
with models of \Hp\ region evolution shows that Sh2-219 is probably 10$^5$~yr old.  
The age and origin of the near-IR cluster observed on the border of 
Sh2-219 remain unknown.}
{}


 \titlerunning{Star formation near Sh2-219}
 \authorrunning{L.~Deharveng, B.~Lefloch et al.}
 
\maketitle

\section{Introduction}

We have previously suggested (Deharveng et al.~\cite{deh03a}) that Sh2-219 is a case 
of sequential star formation. Our main argument was the presence of a deeply 
embedded cluster at the periphery of the \Hp\ region, just beyond the 
ionization front. However, the molecular environment of Sh2-219 was unknown. Here we 
present high-angular resolution molecular observations, as well as new near-infrared 
(near-IR) and radio continuum observations, followed by a discussion of the morphology of the whole region 
in the light of these new data.\\

Sh2-219 is an optically-visible \Hp\ region, of diameter 4.4~pc, lying at a distance of 
$5.0 \pm 0.8$~kpc (Deharveng et al.~\cite{deh03a}), in the direction 
$l=159\fdg355$, $b=2\fdg592$. It is a nearly  
perfect Strömgren sphere around a central O9.5V exciting star (see Sect.~5). The mass of 
the ionized gas is 39\msol\ (Leahy~\cite{lea97}). Sh2-219 is surrounded by a 
thick ring of atomic hydrogen, except for a gap in the southwest. 
This ring has a FWHM of 2--3~pc, a mean density of 
9~atoms~cm$^{-3}$, and a mass of 97\msol\ (Roger \& Leahy~\cite{rog93}). We have previously reported 
(Deharveng et al.~\cite{deh03a}) the detection in the near- and mid-IR of a cluster   
southwest of Sh2-219, at the border of the ionized region, and in the direction of 
the \HI\ gap. This cluster contains highly-reddened objects ($A_V$ up to 14~mag for objects 
detected in all three $JHK$ bands). It also contains an H$\alpha$ 
emission-line star (no.~139 in Deharveng et al.~\cite{deh03a}) presenting a near-IR excess, 
and an ultracompact (UC) \Hp\ region (Leahy~\cite{lea97}). The coordinates of these 
objects are given in Table~1.

Figure~1 shows Sh2-219 as it appears in the optical and the near infrared.

\begin{figure*}
\includegraphics[width=185mm,angle=0]{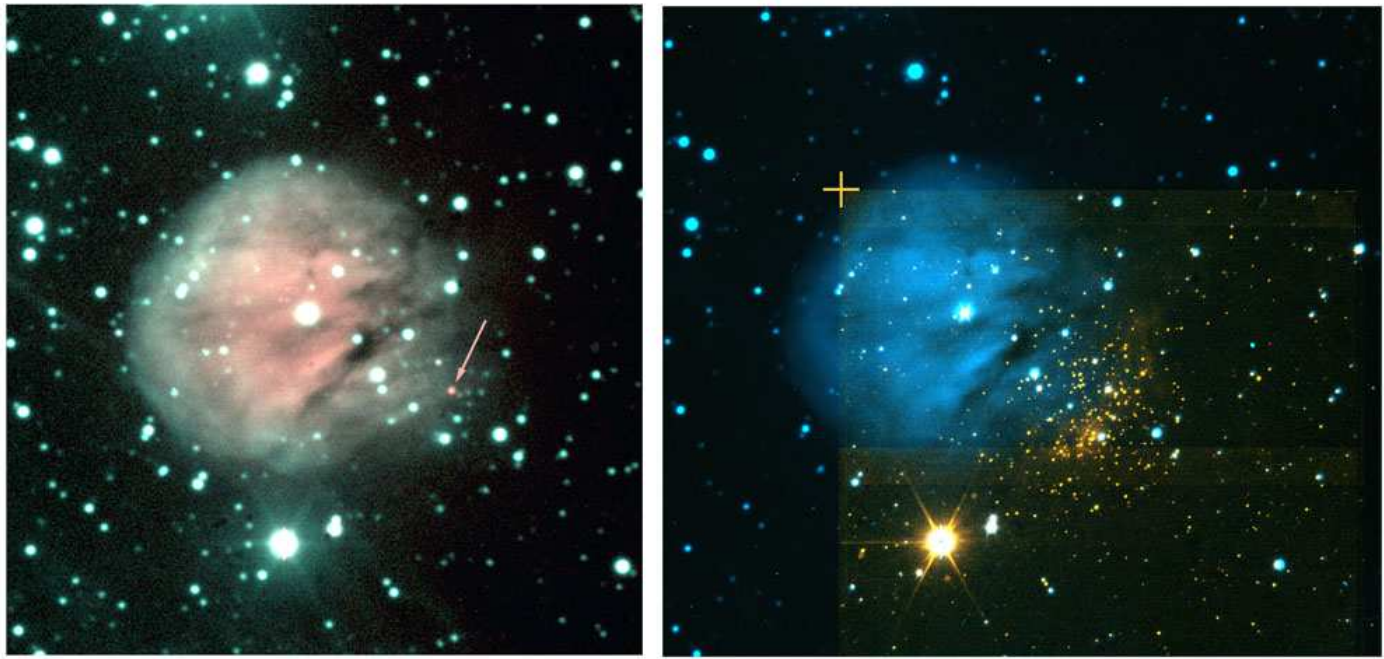}
\caption{Sh2-219. {\it Left: }composite colour image of Sh2-219 in the optical. 
North is up and east is left. The size of the field is $5\farcm1 \times 5\farcm2$.
Pink corresponds to the H$\alpha$ 6563\AA\ emission, and turquoise to the [SII] 
6717--6731\AA\ emission, enhanced near the ionization front. The arrow points 
to the H$\alpha$ emission-line star (no.~139 in Deharveng et al.~\cite{deh03a}). 
{\it Right: }composite colour image of Sh2-219 in the optical and the near-IR. Blue 
corresponds to the H$\alpha$ emission of the ionized gas, 
and orange to the $K'$ emission of the stars (new $K'$ observations discussed in Sect.~4); 
the cross shows the northeastern limits of the $K'$ frame.}
\end{figure*}

\begin{table}[h]
\caption{Coordinates of the objects discussed in the text}
\begin{tabular}{lllllll}
 \hline\hline
 Object & \multicolumn{3}{c}{RA~(2000)} & \multicolumn{3}{c}{Dec~(2000)}\\
 & h & m & s & $\degr$ & $\arcmin$ & $\arcsec$ \\
  \hline
 Exciting star of Sh2-219$^{\mathrm{a}}$  &  4 & 56 & 10.61 & 
+47 & 23 & 35.4 \\
 H$\alpha$ emission star no.~139$^{\mathrm{a}}$  &  4 & 56 & 03.76  & +47    &22  &58.2      \\
 UC \Hp\ region$^{\mathrm{b}}$                 &    4  & 56    & 02.1    &   +47  &23    &07.0  \\ 
 \hline \\
\end{tabular}

$^{\mathrm{a}}$ Deharveng et al.~(2003a)\\
$^{\mathrm{b}}$ This paper
\end{table}

\section{Observations}
\subsection{CO observations}

In December 2002 we observed the emission of the molecular gas associated with 
Sh2-219, in the \twco~\jtwotoone\ line ($\nu$=230~GHz), using the IRAM 30-m 
telescope (Pico Veleta, Spain). 
We  mapped an area of $14\arcmin \times 15\arcmin$ with the HERA nine-channel 
heterodyne array (Schuster et al.~\cite{sch04}). 
The beam size of the telescope is $11\farcs7$ at this frequency. The 
elements of the array are arranged in a $3\times 3$ matrix, with a separation of 
$24\arcsec$ on the sky between adjacent elements.
The data were acquired by drifting the telescope in right ascension 
using the standard `on-the-fly' 
technique. A reference position was taken 10$\arcmin$ east of the H$\alpha$ star. 
It was checked to be free of emission down to 0.15~K in antenna temperature 
($1\sigma$, in a velocity interval of 
$0.25\kms$) between $-70$ and $-6\kms$. The beam pattern on the sky was 
rotated by 18.5 degrees with respect to the right 
ascension axis by means of a K-mirror mounted between the Nasmyth focal plane and 
the cryostat of the heterodyne array. When drifting the telescope 
in right ascension, two adjacent rows are separated by $7\arcsec$, which results in 
a slightly undersampled map in declination (the Nyquist sampling step is $5\farcs5$).  
Details about the HERA array and the K-mirror can be found at 
http://www.iram.fr/IRAMFR/PV/veleta.htm. 

A digital autocorrelator with a spectral resolution of 78~kHz was used as a 
spectrometer; the resolution was degraded later to obtain a velocity resolution 
of $0.25\kms$ at 1.3~mm. The observing conditions 
were typical for the time of the year, with typical 
system temperatures of 400 to 500~K. Pointing was checked every 90 minutes by 
scanning across nearby quasars; it was found to be stable to better than $3\arcsec$.  
 
Supplementary observations in the \ceio~\jtwotoone\ and \jonetozero\ transitions 
at 219.560319~GHz and 109.782182~GHz, respectively,  
were carried out at two positions in the nebula, using the `standard' 
heterodyne receivers at the IRAM 30-m telescope. The weather conditions were very 
good, with system temperatures of 145~K and 260~K at 3~mm and 1.3~mm respectively. 

Given the extended size of the CO-emitting region, the antenna temperature  
is a satisfactory approximation to the CO line brightness. On the other hand, for the 
\ceio\ emission,  which is much more compact, the main-beam brightness temperature scale 
$T_{\rm mb}$ is a reasonable approximation to the intrinsic line brightness. We adopt 
values of 0.70 and 0.50 for the main-beam efficiency at the frequencies of the 
\ceio\ \jonetozero\ and \jtwotoone\ transitions respectively.

In all the maps presented hereafter, the coordinates are expressed in arcsecond offsets 
with respect to the position of the H$\alpha$ emission-line star (Table~1).

\subsection{Other molecular observations}

As part of a larger observational program, we observed Sh2-219 with
the Very Large Array on 2005 November 16.  We searched for emission
from methanol at 44~GHz, water at 22~GHz, and ammonia (in the (1,1)
and (2,2) lines) at 23~GHz.  The array was in its most compact
configuration, providing spatial resolutions of 1$\farcs$5 in the 7~mm
band and 3$\farcs$3 in the 1.3~cm band. The pointing center was 
RA~(2000)$=4^{\rm h}~56^{\rm m}~03\fs3$, Dec~(2000)$=+47^{\circ}~22\arcmin~57\arcsec$, 
and the primary beam was 1$\arcmin$ and 2$\arcmin$, respectively in the 
7~mm and 1.3~cm band. Further observational details
will be found in Deharveng et al.~(2007, in preparation). With 4$\sigma$  
detection limits of 45, 23, and 8 mJy~beam$^{-1}$ for the methanol, water, 
and ammonia lines, respectively, we failed to detect any of these molecular 
transitions.

\subsection{Near-IR observations}

Near-IR observations were carried out through
$H$ and $K'$ broad-band filters, and [{\it Fe\,{\sc{ii}}}] and $H_2$ narrow-band filters, with the
NICS camera (Baffa et al.~\cite{baf01}) at the 3.58-m Telescopio 
Nazionale Galileo (TNG, Canary Islands) 
on January 4, 2004. The plate scale is $0\farcs25$/pixel,
yielding a field of view of $4\farcm2 \times 4\farcm2$.
The $H_2$ and [{\it Fe\,{\sc{ii}}}] filters are centred at 2.12~$\mu$m and 1.67~$\mu$m   
respectively. Four pairs of on-source, off-source images on a 
dithered pattern were acquired through the
$H_{2}$ and [{\it Fe\,{\sc{ii}}}] filters (total on-source integration time 10 minutes).
Two series of four dithered on-source and off-source images 
were acquired through the $K'$ filter (total on-source 
integration time 80 sec) and one series of on-source and off-source images 
through the $H$ filter (total on-source integration time 80 sec). 
Flat-field images were obtained from sky observations at sunset through all
four filters.

Each frame was first corrected for crosstalk using
the routine provided on the TNG web page 
(http://www.tng.iac.es). Data reduction was then performed
by using standard IRAF routines. After flat-fielding all
frames, the four off-source images of each series were median-combined and subtracted
from each on-source image of the same series in order to
remove the background emission. The sky-subtracted
images thus obtained were corrected for distortion and bad pixels, 
and were averaged together.

Aperture photometry was done using the DAOPHOT (Stetson~\cite{ste87}) package
in IRAF. We selected an aperture $\sim 1$ FWHM 
in radius and the sky intensity was estimated from its median value 
in a $\sim 1$-FWHM wide annulus, between 2 FWHM and 3 FWHM from the aperture centre 
(FWHM $\sim0\farcs8$ in $H$ and $\sim0\farcs7$ in $K'$) . 
Stars were found using DAOFIND and 
the list was then corrected for false or missed detections after a visual 
check of the images.

Although the seeing was good, the night
was barely photometric, so we calibrated our $H$ and $K'$ photometry
by using that performed on the same field by Deharveng et al.~(\cite{deh03a}). 
We calculated the difference 
in magnitudes for each coinciding source and plotted it against 
our derived instrumental $H - K'$. A colour term was present 
in $K'$because of the difference of the $K$ and $K'$ filters.
After a best-fit calibration, an rms residual of $\sim 0.2$ mag was 
present in the difference between our photometry and that of 
Deharveng et al.~(\cite{deh03a}),
probably due to the different seeing, to the fact that we are comparing aperture 
and PSF-fitting photometry, and also possibly to the intrinsic
variability of young stars. Our resulting photometry is in the same $HK$
system as Deharveng et al.~(\cite{deh03a}). Sources brighter than $\sim$11~mag in 
$H$ or $K$ are in the non-linear range of the detector; in these cases, we adopted
the measurements by Deharveng et al.~(\cite{deh03a}).  The limiting detection magnitudes
are $\sim18$ in $K$ and $\sim18.5$ in $H$, much deeper than 
obtained by Deharveng et al.~(\cite{deh03a}). We estimate 
a completeness limit of $K\sim16.5$ by plotting the number of
detected sources in 1-mag bins. 
A total of 616 near-IR sources with detection both in the $H$ and in the $K$ band and
53 with detection in only one band were found over the field.

As for the narrow-band images, we removed the continuum contribution falling
in the band by scaling both the narrow-band and the corresponding broad-band
frame ($K'$ for $H_{2}$ and $H$ for [{\it Fe\,{\sc{ii}}}] according to the signals measured
for a few stars in the field (which are line-free, pure-continuum sources)
and subtracting one from the other. The resulting frame contains only line emission.

\subsection{Radio continuum observations}

A radio map of Sh2-219 with a resolution of 40$\arcsec$, obtained at 20~cm 
with the VLA D array by Fich~(\cite{fic93}), showed a single 
radio-continuum source. A later 20~cm map, obtained by  Leahy~(\cite{lea97}) 
with the VLA C array,  
had better resolution and showed the presence of an unresolved source south-west 
of Sh2-219. But no details were given in Leahy's paper about this ultracompact 
component.

We used archival VLA data in order to assess the
radio properties of both Sh2-219 and the UC \Hp\ region to the south-west.
The observations were
made at 20~cm and 6~cm in the D and C configurations (programme AL216)
and at 20~cm in the B configuration (programme AF346).  The B array
20~cm data were consistent with the C array 20~cm data, and for simplicity 
will not be discussed further.

The AL216 observations were made in January 1991 (C array) and April
1991 (D array). For both observation dates the primary calibrator was
3C48 and the secondary calibrator was 0435+487.  The adopted flux
densities for 3C48 are 15.41 Jy at 20 cm (1.46~GHz) and 5.16 Jy at
6 cm (4.86~GHz). The boot-strapped
flux densities for 0435+487 were 1.25 and 0.52 Jy (at 1.46 and 4.86
GHz, respectively) in January, and 1.24 and 0.51 Jy in April.  Data
reduction followed standard procedures for VLA continuum data.

The data obtained in the C configuration at 20~cm provide a 
uv-coverage similar to those obtained in the D configuration at 6~cm; hence
these are the appropriate datasets for comparing the flux densities at
the two frequencies, for both Sh2-219 and the UC \Hp\ region.
The C configuration's 6~cm data provide the highest angular resolution
and hence give a stronger limit on the size of the UC \Hp\ region.
The angular resolutions and noise levels are $14\arcsec
\times 12\arcsec$ and 140~$\mu$Jy~beam$^{-1}$ (C configuration, 20~cm),
$15\arcsec \times 11\arcsec$ and 65~$\mu$Jy~beam$^{-1}$ (D configuration, 6~cm),
and $4\arcsec \times 3\arcsec$ and 60~$\mu$Jy~beam$^{-1}$ (C configuration, 6~cm).
The D configuration 6~cm image is shown in Fig. 2.

In addition, a 1.3~cm (23.71~GHz) continuum image was produced from
the ammonia observations mentioned in Sect. 2.2.  This image has a
resolution of $2\farcs6 \times 2\farcs3$ and a noise level of 0.2~mJy~beam$^{-1}$.

%
\begin{figure}
\includegraphics[angle=0,width=85mm]{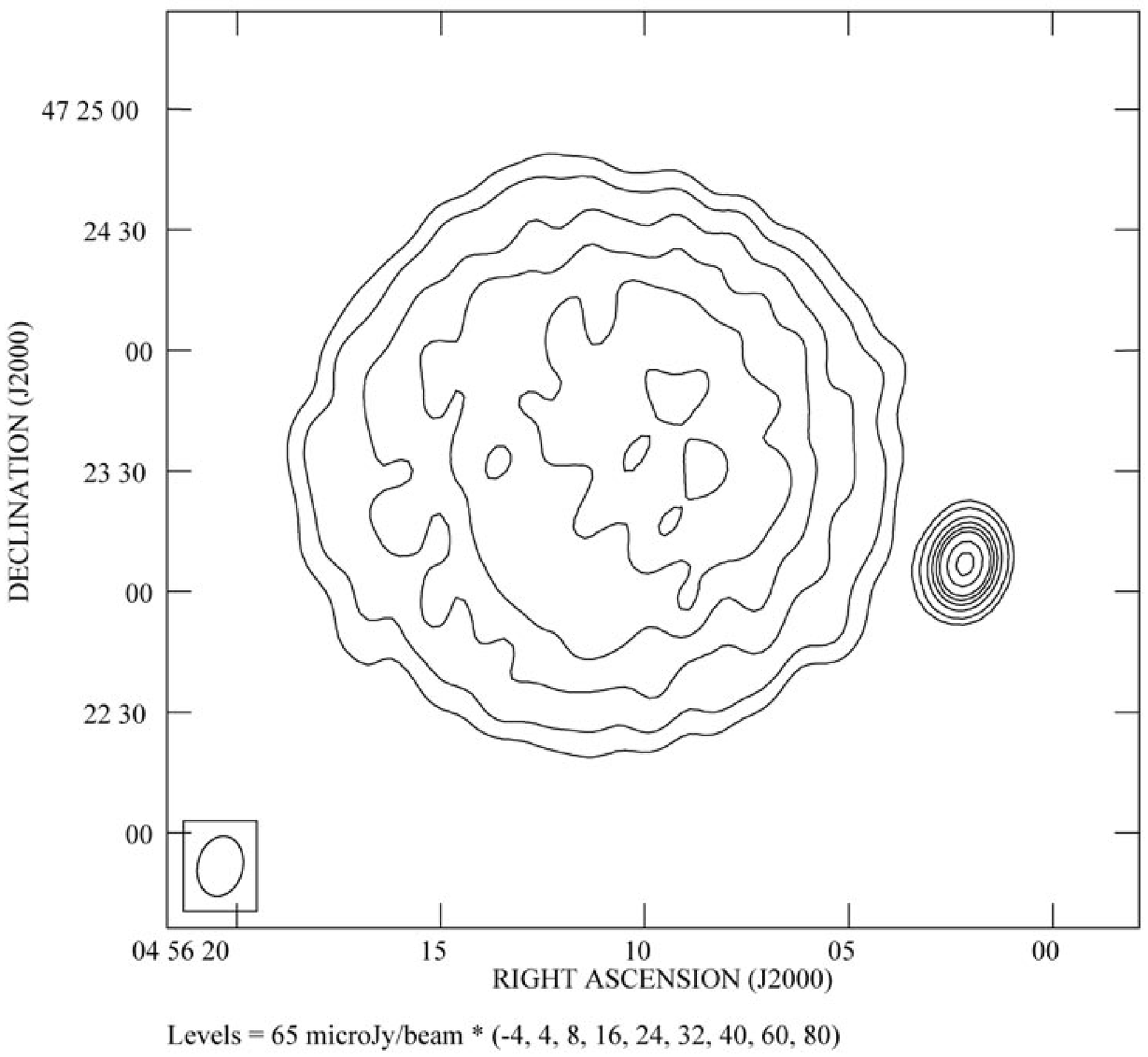}
   \caption{ Radio continuum image obtained with the VLA at 6~cm.  The
angular resolution of the image (indicated in the lower left corner)
is $15\arcsec \times 11\arcsec$. The contour levels are multiples of
the 65~$\mu$Jy~beam$^{-1}$ image rms.}
\end{figure}
%

\section{The molecular gas}
\subsection{ Distribution and kinematics}

\begin{figure*}
\includegraphics[width=180mm,angle=0]{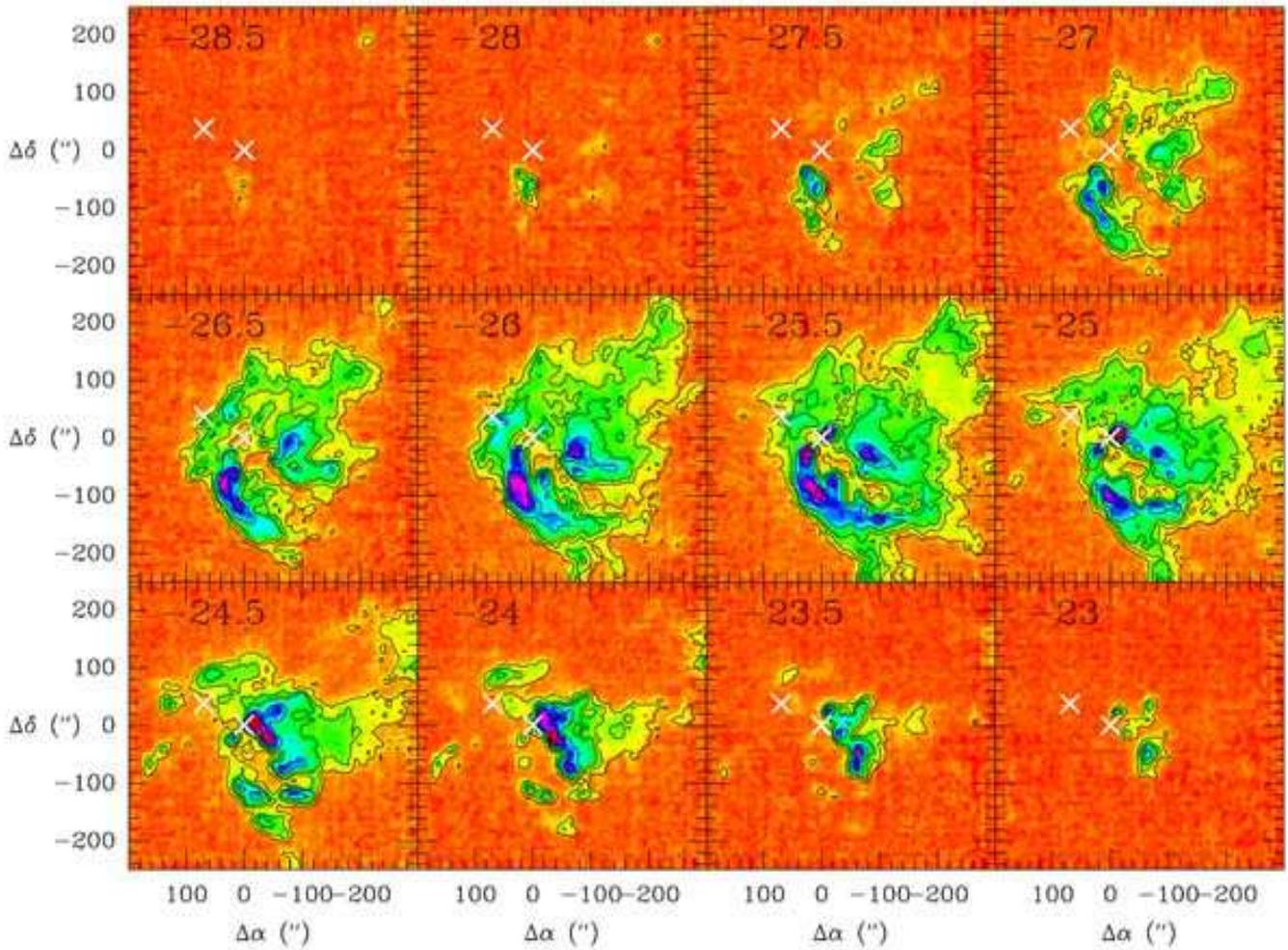}
\caption{Channel maps of the CO $\jtwotoone$ emission as observed with HERA and 
integrated in velocity intervals of $0.5\kms$; the central velocity is marked in 
the upper left corner of each panel. The contour levels are 1, 2, 3, 5, \dots, 
13~K$\kms$. The positions of the exciting star of Sh2-219 and of the H$\alpha$ 
emission-line star (at 0,0) are marked with white crosses.}
\end{figure*}

The distribution of integrated CO emission as a function of velocity is shown in Fig.~3.  
CO emission  is detected to the southwest of Sh2-219 as a circular cloud of 
diameter $\sim 12\pc$, 
adjacent to the \Hp\ region; the CO lines have typical antenna 
temperatures of 6--10~K. There is no molecular emission observed in the direction 
of the \HI\ ring, nor is any CO associated with a possible photodissociation region   
surrounding the ionized gas. 

The overall CO emission covers a velocity range of $\sim 6~\kms$. The molecular 
emission associated with Sh2-219 is particularly bright in the range $-26.5\kms$ 
to $-25.0\kms$, where the bulk of the emission is detected; the systemic 
velocity of the cloud is $\sim -26\kms$. This is in 
good agreement with previous velocity measurements 
($V({\rm CO})=-24.5~{\rm km~s}^{-1}$, 
Blitz et al.~\cite{bli82}; $V({\rm CO})=-25.0~{\rm km~s}^{-1}$, Wouterloot \& 
Brand~\cite{wou89}). No emission is detected at $-31.0~{\rm km~s}^{-1}$, the velocity 
measured for the ionized gas by Fich et al.~(\cite{fic90}). Note however that 
a mean velocity of $-24.6~{\rm km~s}^{-1}$ was measured by Georgelin \& 
Georgelin~(\cite{geo70}) for Sh2-219, in good agreement with that 
of the adjacent molecular cloud. 

The CO lines are relatively narrow, about $1.5\kms$. There is no 
evidence for a bipolar outflow which could be the signature of ongoing star 
formation in the cloud. However, detection of compact outflows, typical of 
UC \Hp\ regions, may be hampered by the spatial undersampling of the data 
and the low spatial resolution (one telescope beam corresponds to $0.3\pc$ at a distance 
of 5 kpc).

Figure~4 shows the overall distribution of the ionized gas, of the atomic 
hydrogen, and of the molecular gas. These components are clearly separated. 
{\it Two maxima of CO emission are found at the ionization 
front}, at offset positions $(-19\arcsec,+9\arcsec)$ and $(+23\arcsec,-32\arcsec)$. 
Figure~5 shows that these condensations are elongated along the ionization front.
The former peak coincides with the UC \Hp\ region (see Sect.~5); 
this condensation has a beam-deconvolved FWHM of 
$17\arcsec$ ($0.4\pc$). The 
latter peak coincides with some highly-reddened stars in the near-IR cluster. Its  
beam-deconvolved FWHM is similar (0.5~pc). Also, the CO emission map 
{\it suggests the presence of a 
`chimney' in the molecular cloud which nearly splits the cloud in two}. 
This chimney is oriented southwest to northeast (approximate projection angle of 
$-135^{\circ}$). It is best detected in the blueshifted and redshifted gas 
(see e.g.\ the panels at $-27$ and $-24\kms$ in Fig.~3), but is 
nonetheless present in the ambient gas at $-26\kms$. The H$\alpha$ emission-line star 
lies near the head of the chimney, and close to its axis. This 
chimney has an approximate length of 5~pc and a width of 0.5--1.0~pc.

%
\begin{figure}
 \includegraphics[width=85mm]{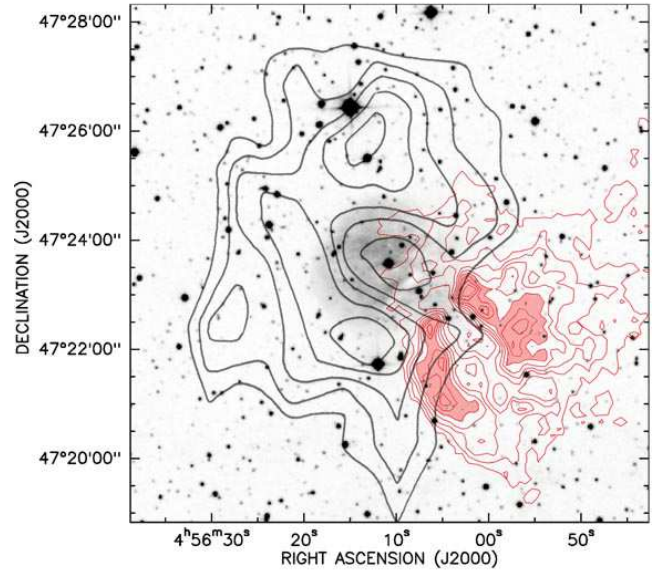}
  \caption{The ionized, atomic and molecular components of the Sh2-219 complex. The 
  \Hp\ region (DSS-2 red survey) appears as a grey scale. The black contours show 
  the ring of atomic material surrounding the ionized gas (from Roger \& Leahy  
  \cite{rog93}). The thin red contours are for the velocity-integrated CO emission. 
  The first contour is $5.8\K\kms$ and the contour interval is $2.9\K\kms$.}
\end{figure}

%
\begin{figure*}
\includegraphics[width=185mm,angle=0]{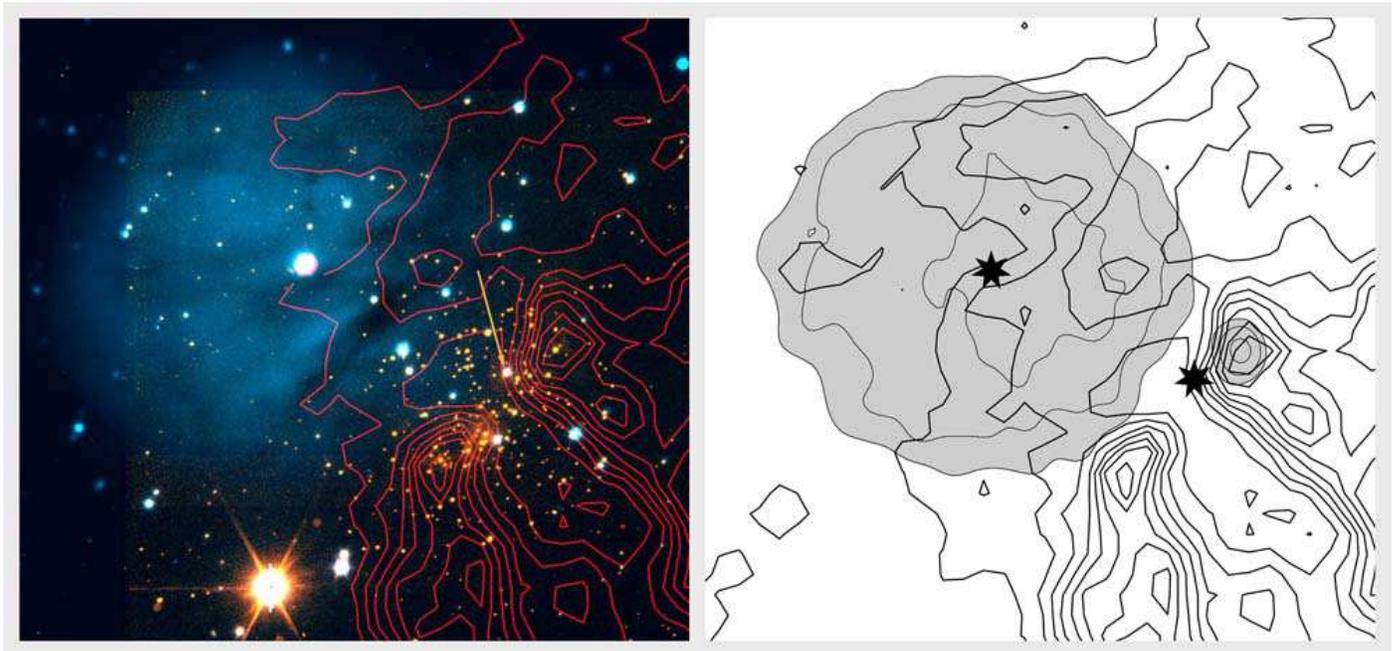}
\caption{{\it Left:} Composite colour image of Sh2-219. The H$\alpha$ emission of the ionized gas
appears in blue, the stellar $K'$ emission in orange. The red contours are for the 
velocity-integrated temperature distribution of the CO emission. Note how the CO maxima 
are elongated along the ionization front. The arrow points to the H$\alpha$ emission-line 
star. {\it Right:} The CO contours are superimposed to the radio continuum emission 
map at 20~cm.}
\end{figure*}

To gain more insight into the kinematics of the cloud, we have studied the profile 
of the CO emission along a southeast-to-northwest cut across the chimney, 
i.e.\ perpendicular to its main axis. Figure~6 shows the CO line intensity along the cut, 
as a function of the velocity. The gas appears to possess a small velocity 
gradient of $0.17\kms\pcmu$ from $\rm Y= -100\arcsec$ to 
$\rm Y= +350\arcsec$. The cavity in the gas is visible as the region 
of low emissivity near $\rm Y= 0\arcsec$. Two components, with velocities 
shifted with respect to that of the ambient gas, are detected near 
this position, one on the northern side at $-24.0\kms$ and one on the southern side 
at $-27.5\kms$ (Fig.~6). These components form the walls of the cavity. The southern  
wall reaches an extreme velocity of $-31\kms$, and the northern wall of $-22\kms$. 
This is quite puzzling: it may indicate that 
the inner walls of the chimney are in rotation around 
the chimney axis, with the northern wall receding from us and the southern 
wall approaching us.

Faint CO emission at $\sim 6~\kms$ is observed south of \mbox{Sh2-219}  
($\Delta \delta < 250 \arcsec$). This is most probably  
foreground emission, unrelated to this region.

%
\begin{figure*}
\includegraphics[width=185mm,angle=0]{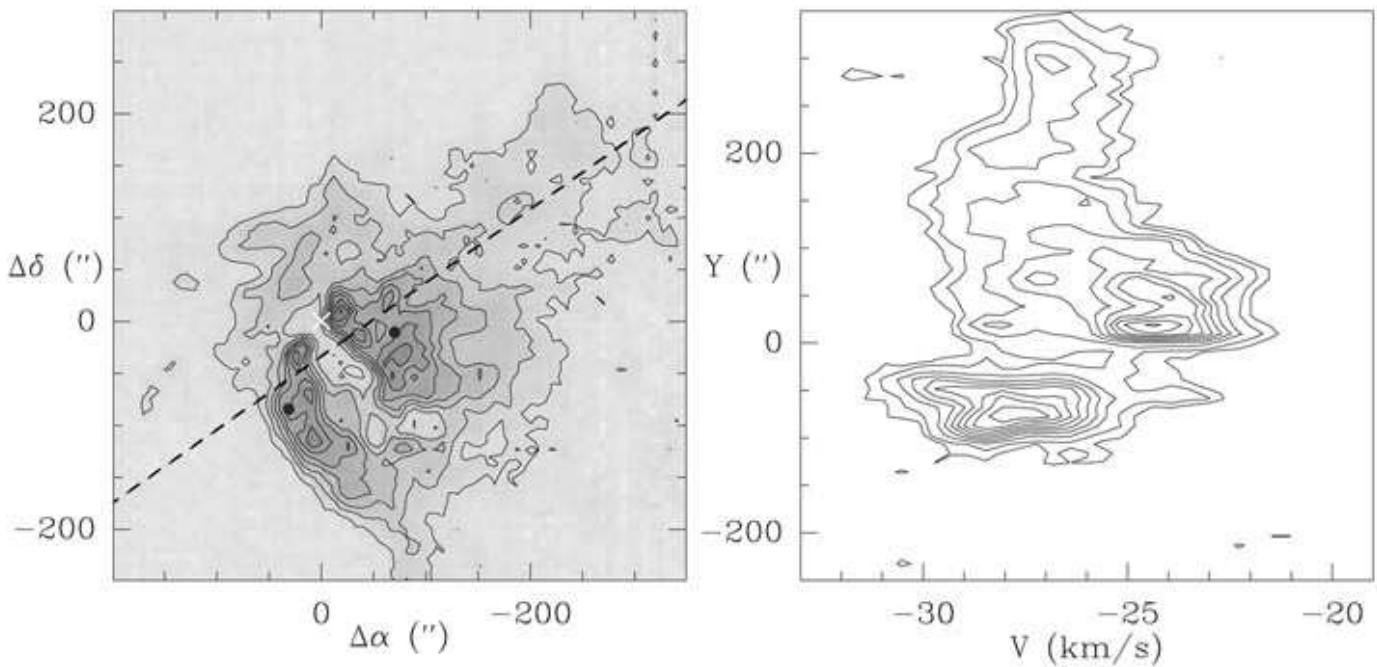}
\caption{
{\em Left:}~Distribution of the CO emission integrated between $-28.0\kms$ and 
$-22.5\kms$; contours are 0.5, 1, 2, 3 \dots times 5.75~K~km~s$^{-1}$. The position of the 
H$\alpha$ star at ($0\arcsec$, $0\arcsec$) is marked with a white cross. Coordinates are 
arcsecond offsets with respect to the H$\alpha$ star. The positions of the two 
\ceio\ observations are indicated by black dots.  
{\em Right:}~Velocity-position diagram along a 
cut perpendicular to the chimney's main axis. The cut was made $28\arcsec$ southwest 
of the H$\alpha$ emission star, along the south-east to north-west direction shown as a 
dashed  line in the left panel. Y=0 corresponds to the projection of the H$\alpha$ star 
on the cut.
}
\end{figure*}

\subsection{Physical conditions}

The brightest CO emission is detected at the head of the chimney, in the two 
condensations adjacent to the ionization front, with antenna 
temperatures of up to 12~K, indicating a gas kinetic temperature 
of $\sim15$~K. The CO antenna temperature along the chimney is lower, typically 8--10~K.  

The molecular material associated with Sh2-219 has been mapped only in 
$^{12}$CO~(2--1), so for the mass determination we rely on the
empirical relation between cloud mass and integrated emission $\int T({\rm CO})dv$. We use  
the CO to H$_2$ conversion factor 
$X=1.9~\times~10^{20}$~mol~cm$^{-2}$~(K~km~s$^{-1}$)$^{-1}$, 
as determined by Strong \& Mattox
(\cite{str96}). Including all emission above 1~K~km~s$^{-1}$ ($\sim
3.5\sigma$) this yields $M_{\rm CO}=2100~M_{\odot}$. The
equivalent radius of the cloud above that emission level is 
$r_{\rm e}=\sqrt{{\rm Area}/\pi}=5.8$~pc. 
The linewidth of the composite cloud spectrum is $\Delta~V=2.4$~km~s$^{-1}$, 
but this reflects the fact that different parts of
the cloud have different velocities (cf. Fig.~3), rather than turbulence. 
Using the typical profile
width (1.5~km~s$^{-1}$; see Sect. 3.1) instead, and assuming a density
profile $\propto r^{-2}$, we find a virial mass 
$M_{\rm vir}=126~\times~r_{\rm e}~(\Delta~V)^2=1800$\msol, in good agreement 
with the CO mass derived above, and consistent with the cloud as a whole being in 
gravitational equilibrium.

Likewise, the mass of the gas associated with the chimney, corresponding to  
$\int~T({\rm ^{12}CO})dv~\ge~10$~K~km~s$^{-1}$, is 770\msol; the
equivalent radius of this emission region is about 2~pc (only an indicative
value of course, as the chimney emission region is elongated). The
composite CO spectrum of the chimney emission has an equivalent width
of $\sim~3.1$~km~s$^{-1}$ - wider than that of the cloud as a whole, because
the chimney gas shows the largest range in velocity (see Figs~3 and 6). Here also, 
the individual profiles have a width of about 1.5~km~s$^{-1}$.\\

Two positions in the cloud, indicated in Fig.~6, were observed in the 
$\ceio~\jtwotoone$ and $\jonetozero$ lines. The 
lines are narrow (1~km~s$^{-1}$) and weak: typically $T_{\rm mb}=0.2$~K. 
At offset position ($-70\arcsec,-10\arcsec$) -- behind the northern border
of the chimney -- we find $T_{\rm mb}^{21}=0.36$~K and $T_{\rm mb}^{10}=0.22$~K. At this 
position, the CO \jtwotoone\ antenna temperature is $\approx~7\K$. We have estimated 
the physical conditions from a simple calculation in the large-velocity gradient 
approximation, assuming uniform density and 
temperature in the emitting region. Adopting a kinetic temperature of 15~K and a linewidth 
of 1.0~km~s$^{-1}$, we obtain 
$n(\htwo)=8.0\times 10^3\cmmt$ and $N(\ceio)=2\times 10^{14}\cmmd$. Adopting 
a standard \ceio\ abundance of $2\times 10^{-7}$ (Frerking et al.~\cite{fre82}), 
this corresponds to an \htwo\ gas column density $N(\htwo)$ of $1.0\times 10^{21}\cmmd$. 
The \ceio\ lines are optically thin with $\tau^{21}=0.054$ and $\tau^{10}=0.015$. 
The excitation temperature of the \jtwotoone\ transition is only 11~K, i.e.\ it is not
thermalized. On the other hand, the \jonetozero\ transition is close to thermalization. 
We also observed the \ceio\ emission at the offset position ($+30\arcsec,-84\arcsec$), near 
the southern border of the cloud. The lines intensities are comparable and we derived 
a similar density and a similar column density in the gas traced by \ceio . 

If we assume that the emission detected at the two 
observed positions is representative of the bulk of the chimney gas, 
the derived $N({\rm H}_2)$, together with the 2~pc radius found from the 
$^{12}$CO-observations, leads to a mass of 270~$M_{\odot}$ for the 
chimney gas. This (very uncertain) value is almost three times less than 
what is derived from the $^{12}$CO-emission.  
If the two CO clumps are density enhancements (as opposed to regions with an 
elevated temperature), they would contain much more material than the walls 
of the chimney and would consequently exhibit stronger C$^{18}$O lines, 
which would lead to a higher mass estimate. 
As no C$^{18}$O observations were carried out towards the two clumps this 
cannot be confirmed.

\section{The stellar population of the cluster}

\begin{figure}
   \centering
   \includegraphics[angle=0,width=85mm]{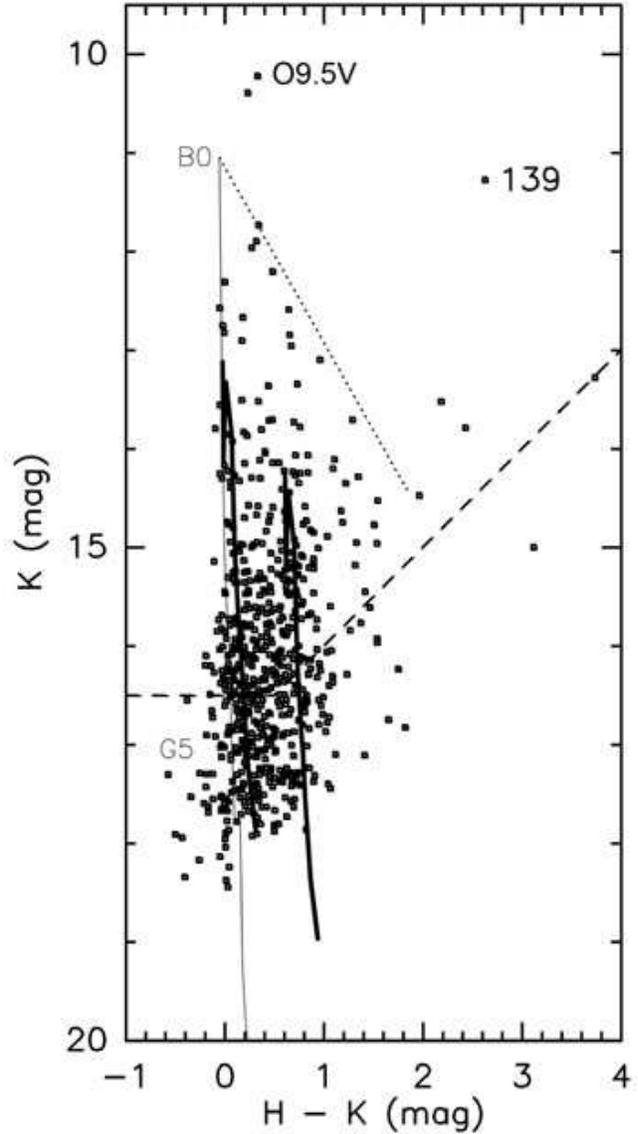}
   \caption{$K$ vs.\ $H-K$ for all sources with
   detection in both bands. The dashed lines mark the
   completeness limit, whereas the thin line is the locus
   of the ZAMS, for $A_{V}=0$ and a distance of 5~kpc; the absolute magnitudes are from Allen~(\cite{all76}),
   and the colours from Koornneef~(\cite{koo83}). The thick lines are isochrones 
   for pre-main-sequence stars 1 Myr in age (with $A_{V}=0$ and 10 mag, 
   left to right) with masses from 0.1 to 6 $M_{\sun}$ (from 
   Palla \& Stahler~\cite{ps99}). The length of the dotted reddening line  
   originating from a B0 star corresponds to a visual extinction 
   of 30~mag; the adopted extinction law is
   from Rieke \& Lebofsky~(\cite{rie85}). The positions of the exciting star of Sh2-219 (O9.5V) and 
   of the H$\alpha$ emission star (no.~139) are given.
  \label{k-hk:fig}}
   \end{figure}

A colour-magnitude ($K$ vs.\ $H-K$) diagram for all sources 
detected in both bands is shown in Fig.~\ref{k-hk:fig}.
Most of the stars fall in a strip within $A_{V}\sim10$ mag from the
ZAMS, but a few sources exhibit  
larger extinctions (and, possibly, near-IR excesses).
This confirms the results of Deharveng et al.~(\cite{deh03a}): their
colour-colour diagram ($J-H$ vs.\ $H-K$; see their fig.~4) shows
that most of the sources are consistent with reddened main sequence
stars (with $A_{V}\leq15$~mag), whereas only a few exhibit a near-IR excess.
However, our present data unveil the presence of a larger number of near-IR
sources with $H-K>1$~mag than are seen in Deharveng et al.~(\cite{deh03a}). 
These may be the youngest members
of the cluster. Note that the brightest of these reddened sources is 
star no.~139 of Deharveng et al.~(\cite{deh03a}) -- the H$\alpha$ emission-line star -- 
which is $\sim$0.6~mag brighter in $K$ than given by their photometry. 
This may indicate that this source is variable. 

To better examine the cluster morphology, we have mapped the
stellar surface density by counting all sources (above the completeness
limit) in a $20\arcsec \times 20\arcsec$ window with sampling every $10\arcsec$
both east-west and north-south. The
result is shown in Fig.~8a. The locations of the sources
are marked either by small filled squares ($H-K\leq1$) or large 
squares ($H-K>1$). The surface density distribution appears slightly
elongated in a direction roughly parallel to
the ionization front of the \Hp\ region. The reddened sources 
(presumably also the youngest stars in the cluster) 
are even more clearly aligned along it. Most of these sources are located 
towards the two most prominent CO clumps, at offsets of 
$(-19\arcsec,+9\arcsec)$ and $(+23\arcsec,-32\arcsec)$. 
  
%
   \begin{figure*}
   \centering
   \includegraphics[width=160mm]{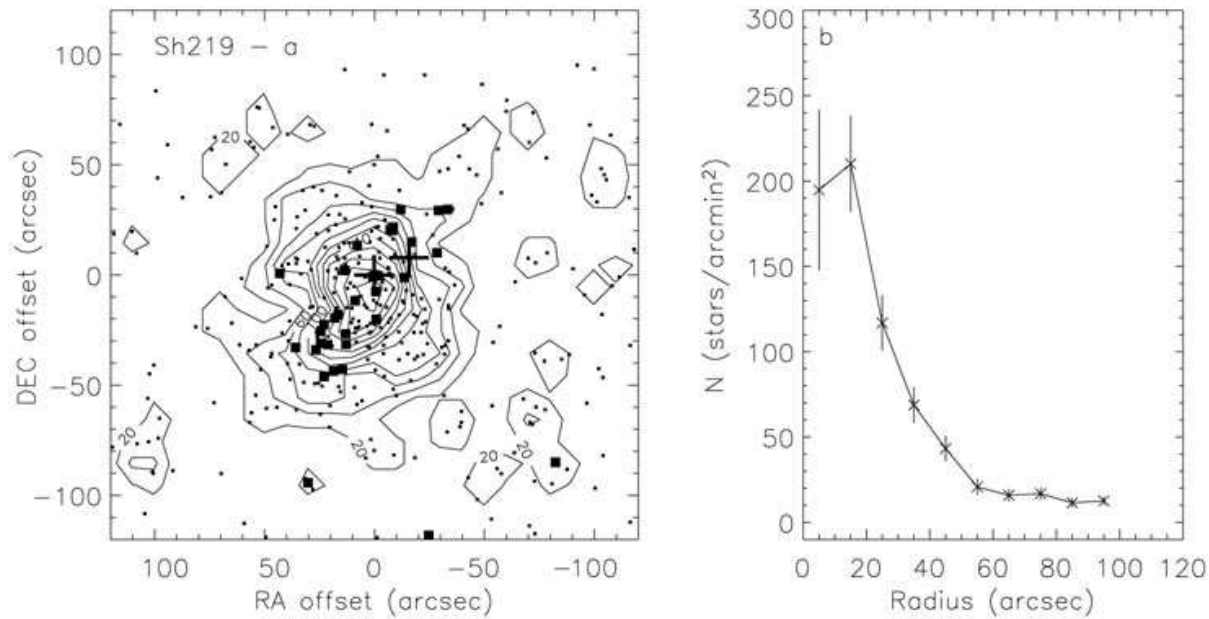}
   \caption{{\bf a} Stellar surface density distribution
   (stars arcmin$^{-2}$); coordinates are arcsec offsets
   from the $H\alpha$ emission-line star no.~139. Small squares mark the positions
   of sources with $H-K \leq 1$, and large squares
   mark the positions of sources with $H-K > 1$. The locations of
   the UC \Hp\ region and of star no.~139 are indicated
   by crosses. {\bf b} Radial stellar density distribution, centred at the surface 
   density peak.}
   \end{figure*}
%

The radial distribution of the stellar surface density can be shown by counting
all sources (above the completeness limit) in $10\arcsec$-wide annuli centred at the
cluster's centre, at RA~(2000)$=4^{\rm h}~56^{\rm m}~03\fs9$, 
Dec~(2000)$=+47^{\circ}~22\arcmin~51\arcsec$. The result is shown in Fig.~8b. 
All derived parameters 
for the cluster are listed in Table~2. From the radial density distribution 
of Fig.~8b, we have estimated the number of cluster members 
(down to the completeness limit) as in Testi et al.~(\cite{tes97}). 
The cluster size is derived from the radial density distribution;  
Table~2 gives the radius at which the surface density falls almost to the 
background value ($\sim$15 stars~arcmin$^{-2}$). 
The completeness limit in mass, $M_{\rm compl}$, can be estimated from 
the completeness limit at $K$ by adopting a set of evolutionary tracks. 
From Palla \& Stahler~(\cite{ps99}), $K = 16.5$ corresponds to a pre-main-sequence star
$10^{6}$ yr old of $0.4~M_{\sun}$ or $10^{7}$ yr old of $1.1~M_{\sun}$. 
Assuming an $A_{V}$ of 10~mag, these mass limits increase to $1.0~M_{\sun}$ 
and $1.4~M_{\sun}$, respectively. Hence the cluster population is probably 
well sampled down to stars of $\sim 1~M_{\sun}$. This means that 
the number of members down to $K=16.5$ is a lower limit to 
the total number of members; assuming a Scalo-like IMF (Scalo~\cite{sca98}), the
correction amounts to a factor between two (for $M_{\rm compl}=0.4$\msol) and six 
(for $M_{\rm compl}$ $=1~M_{\sun}$), without including unresolved
companions and stars below the hydrogen burning limit.

\begin{table}[h]
\caption{Cluster parameters}
\begin{tabular}{ll}
 \hline\hline
Number of members with $K \leq 16.5$ 				& $\sim175 $\\
Fraction of members with $H-K > 1$ 				& 22\% \\
Radius 								& 1.5~pc \\
Expected number of members$^{\mathrm{a}}$ down to 0.1~$M_{\sun}$     	& 380 (1040) \\
Expected number of members$^{\mathrm{a}}$ with $M > 10~M_{\sun}$ 		& 2 (4) \\
Total stellar mass$^{\mathrm{a}}$ 		 		& 280 (750) $M_{\sun}$ \\
\hline\\
\end{tabular}

$^{\mathrm{a}}$ assuming all the stars down to 0.4~(1)~$M_{\sun}$ have been detected

\end{table}

If we knew the cluster age we could compare it with the dynamical age of the 
\Hp\ region. As shown in Fig.~\ref{k-hk:fig}, a simple overlay of pre-main-sequence  
isochrones indicates that the stellar population
is compatible with pre-main-sequence stars $10^{6}$~yr old. However, the cluster age
cannot be constrained by the magnitude-colour diagram, as any
isochrones between zero and several megayears are compatible with it. 
Information about the age comes from the number of sources showing a
near-IR excess. Deharveng et al.~(\cite{deh03a}) found only a few sources
with a near-IR excess in the $J-H$ vs.\ $H-K$ diagram. Even
assuming that all our sources with $H-K > 1$ do exhibit a near-IR excess, this
amounts to only 22\% of cluster members (see Table~2). This shortage of sources
with a near-IR excess may then suggest a cluster where star formation started more than
$10^{6}$~yr ago (Haisch et al.~\cite{hai01}).

As noted above, the distribution of sources with \mbox{$H-K>1$} suggests that most 
recent star formation occurs parallel to the ionization front, possibly in 
the two most prominent CO clumps.  This is confirmed
by the fact that the north-western clump hosts a UC \Hp\ region and that 
the H$\alpha$ emission-line star
lies at its edge. As for the south-eastern clump, our $H_{2}$ image (Fig.~9)
shows  faint elongated line emission in its direction. This could be the signature of a
jet from a young star; we have tentatively identified the driving source with the star  
indicated in Fig.~9 ($K=13.7$, $H-K = 1.3$). But this could be simply fluorescence 
emission from the surface of the molecular cloud, illuminated by UV radiation leaking 
from the \Hp\ region. No other knots of
H$_{2}$ line emission have been found in the field and no \FeII\ line emission has
been detected.

%
   \begin{figure*}
   \includegraphics[angle=-90,width=160mm]{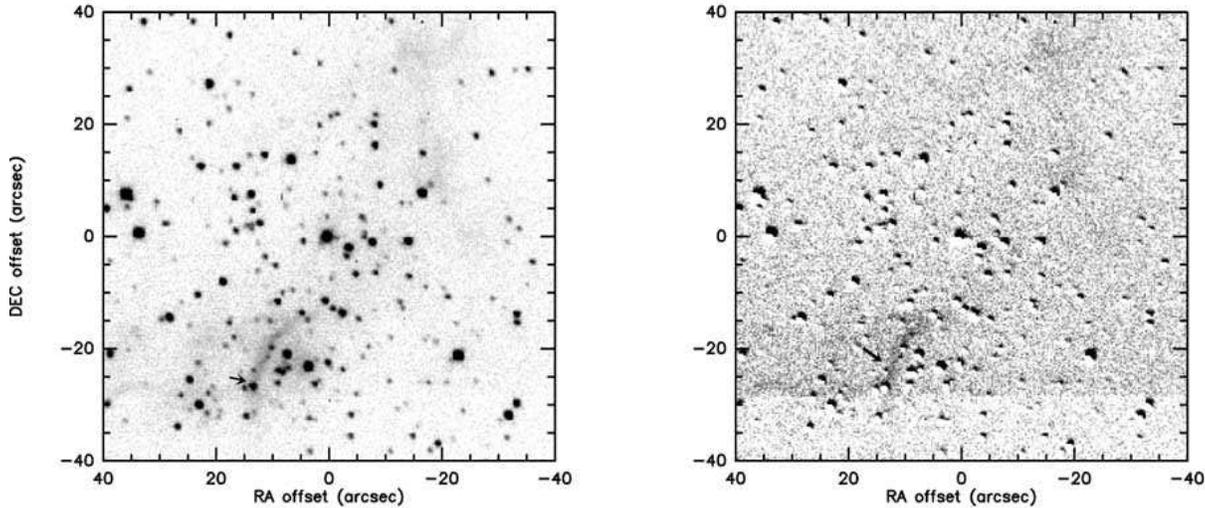}
   \caption{Enlargement of the $H_{2}$ image of the cluster:
   {\it left}, line plus continuum emission within the band,
   {\it right}, after subtraction of the continuum emission. The
   arrows mark the faint emission feature, tentatively 
   identified as a jet (right panel) and
   the possible jet-driving source (left panel). The 
   H$\alpha$ emission-line star lies at (0,0).}
   \end{figure*}

\section{The ultracompact H\,{\scriptsize \bf II} region}

Figure~2 gives the 6~cm radio continuum map of Sh2-219, showing the 
presence of an ultracompact \Hp\ region at its southwestern border. 
Leahy~(\cite{lea97}) showed the 20~cm counterpart of 
this map, i.e. the 20~cm C configuration map of VLA 
program AL216, while we show the 6~cm D configuration map. The focus 
of Leahy~(\cite{lea97}) was the Sh2-219 region; that paper did not discuss 
the UC \Hp\ region to the south-west.

Sh2-219 has a flux density of $140 \pm 7$~mJy at 20~cm  
and of $127 \pm 6$~mJy at 6~cm. These values are consistent with
an optically thin \Hp\ region, with an ionizing photon flux of log
$N_{\rm Lyc}$ = 47.5. According to the calibration of Martins et al.~(\cite{mar05}), 
this flux corresponds to an O9.5V exciting star, in good agreement with 
the $K$ magnitude of the central exciting star of Sh2-219 (Fig.~7).

The UC \Hp\ region has a flux density of $3.2\pm0.3$~mJy at 20~cm,
$6.1\pm 0.4$~mJy at 6~cm, and $4.2\pm0.8$~mJy at 1.3~cm. To within the
uncertainties, these flux densities are consistent with a thermal
spectrum.  The deconvolved source size is $< 0\farcs6 
\times 0\farcs5$, corresponding to $< 15$~mpc$~\times$~12~mpc. The minimum ionizing photon
flux required to maintain ionization of the region is 
log~$N_{\rm Lyc}>46.1$, corresponding to a B0.5V or B1.5V exciting star (according to 
Panagia~\cite{pan73} and Smith et al.~\cite{smi02} respectively). Using the 6~cm
flux density and the upper limit to the size, we derive a lower limit
for the rms electron density in the UC HII region of $1.3 \times
10^4$~cm$^{-3}$.  This electron density is slightly higher than the
H$_2$ density found in the molecular clumps; this is to be expected
as the CO data are averaged over a much larger volume due to the 
lower spatial resolution.

The density and temperature of the molecular clump surrounding the UC 
\Hp\ region are too low to provide a pressure high enough to confine the ionized 
gas. So it seems very unlikely that this UC \Hp\ region is old and has reached  
pressure equilibrium with its surroundings. Furthermore, the radius of the initial 
Strömgren sphere of a star with log~$N_{\rm Lyc}=46.1$ in a medium of density 
$1.3 \times 10^4$~cm$^{-3}$ is about 13~mpc -- almost exactly the 
upper limit we report for the size of the observed UC \Hp\ region. Thus it appears 
that the UC \Hp\ region is quite young, and has not yet had time to 
expand from its initial size. 

\section{Discussion}
\subsection{The H$\alpha$ emission-line star no.~139}

In the following, we try to estimate the luminosity of the H$\alpha$ 
emission-line star no.~139, observed at the southwestern border of 
Sh2-219; this star is identified in Fig.~1. 
This is the brightest $K$ star of the near-IR cluster. 

The integrated H$\alpha$ flux of Sh2-219, from spectrophotometric observations 
(Caplan et al.~\cite{cap00}), is $2.5 \times 10^{-11}~{\rm erg\ s}^{-1}{\rm cm}^{-2}$. 
The H$\alpha$ flux of the emission-line star was determined as follows. First, 
all stars were removed from the H$\alpha$ image using DAOPHOT,  
and the integrated emission of the \Hp\ region was then measured, exempt from stellar 
pollution, in the instrumental units. This allowed us to calibrate the instrumental 
units (ADU$\times$pixels) with respect to flux units 
(${\rm erg\ s}^{-1}{\rm cm}^{-2}$). Next, with DAOPHOT, the stellar fluxes were 
measured in instrumental units from both the (original) H$\alpha$ image and 
the \SII\ image. For all {\em except} the emission-line star the measured signals 
in these two filters were proportional (since they measure continuum emission at 
neighbouring wavelengths but with different filter widths). For the emission-line star, 
the {\em measured} H$\alpha$ minus the H$\alpha$ {\em expected} from the \SII\ 
filter measurement gave us the star's instrumental emission-line signal corrected 
for its stellar continuum. Using our calibration, this difference was then 
converted to an H$\alpha$ emission flux of 
$2.9 \times 10^{-14}~{\rm erg\ s}^{-1}{\rm cm}^{-2}$. At a distance of 5~kpc 
this corresponds to an H$\alpha$ luminosity of $8.6 \times 10^{31}~{\rm erg\ s}^{-1}$ 
($2.3 \times 10^{-2}~L_{\odot}$). 

The H$\alpha$ emission-line star presents a strong near-IR excess which prevents 
us from accurately determining -- and correcting for -- its extinction. Assuming 
that its $J-H$ 
colour is not affected by the near-IR excess, and that its intrinsic $J-H$ 
is 0.0~mag, typical of early type stars, the observed $J-H$ 
of 1.6~mag indicates a visual extinction of $\sim15$~mag. This is close to 
the maximum visual extinction obtained for nearby stars 
in the cluster from measurements in the three ($JHK$) bands 
(Deharveng et al.~\cite{deh03a}). The assumption that star no.~139 has a visual 
extinction of 15~mag leads to an H$\alpha$ luminosity of 
$6 \times 10^{36}$~erg s $^{-1}$ ($\sim1600~L_{\odot}$); this is an  
upper limit, as its $J-H$ colour is probably affected by the near-IR excess. 
A lower limit is obtained assuming a visual extinction of 4~mag, the 
extinction of interstellar origin affecting Sh2-219 (Deharveng at al.~\cite{deh03a}); 
hence an H$\alpha$ luminosity of $1.7\times 10^{33}$~erg~s$^{-1}$ 
($\sim0.45~L_{\odot}$). The H$\alpha$ luminosity of star 139 is 
typical of pre-main sequence objects with mass loss and -- possibly -- outflows 
(Levreault~\cite{lev88}).

The H$\alpha$ star no.~139 is probably a high-mass Herbig Ae/Be star, 
similar to MCW1080 or V645Cyg (which display mass loss and are associated with 
outflows; Levreault~\cite{lev88}). It is also similar to object 25 in 
the AFGL4029 cluster (Deharveng et al.~\cite{deh97}), an emission-line object 
with an H$\alpha$ luminosity of $3.4 \times 10^{-2} L_{\odot}$ (uncorrected 
for extinction), strong extinction and a near-IR excess. 

\subsection{The CO chimney -- H$\alpha$ cavity}

Fig.~10 is a deep H$\alpha$ image of Sh2-219, resulting from the addition of six frames 
(total exposure time of one hour) obtained at the 120-cm telescope of the Observatoire de 
Haute-Provence. Most of the stars have been subtracted using DAOPHOT; the intensity 
of the filaments has been enhanced using Photoshop.  
This H$\alpha$ image shows very thin low-brightness filaments 
south of Sh2-219. Some of these filaments are oriented east-west. They most probably 
belong to the foreground supernova remnant G160.9+2.6, also known as HB9 
(Van den Bergh et al.~\cite{van73}), whose distance is $\sim 1.2$~kpc      
(Leahy \& Roger~\cite{lea91}). Other filaments, indicated by arrows, 
are oriented north-east to south-west and seem to form the walls of a cavity, 
 closing the molecular chimney at 
its south-western extremity. For comparison, Fig.~10 also shows the molecular emission 
superimposed on the H$\alpha$ image. The molecular chimney and the H$\alpha$ cavity 
have the same axis. The length of the cavity is 7.5~pc, from the `head' of the 
chimney in the north-east, near star no.~139, to its southwest extremity. 
Large-scale Herbig-Haro jets or flows, driven by low-luminosity 
sources (Reipurth et al.~\cite{rei97} and references therein), are known to exist. The largest 
of these is HH~111, which has a total extent of 7.7~pc, and is driven by a 25~\lsol\  
class~I source. Also, a few large-scale jets are driven by luminous sources. 
HH~80-81, for example, with a total extent of 5.3~pc, is driven by a $2 \times 10^4$~\lsol\  
young B star (Mart\'{i} et al.~\cite{mar93}). However, none of these jets or flows are 
associated with molecular chimneys or cavities.

The H$\alpha$ emission-line star no.~139, located at the head of the 
chimney and near the axis of the cavity, is well positioned to be at the origin of such 
a structure. Near-IR spectra of this object are needed to ascertain its nature.

%
\begin{figure}
  \includegraphics[width=85mm]{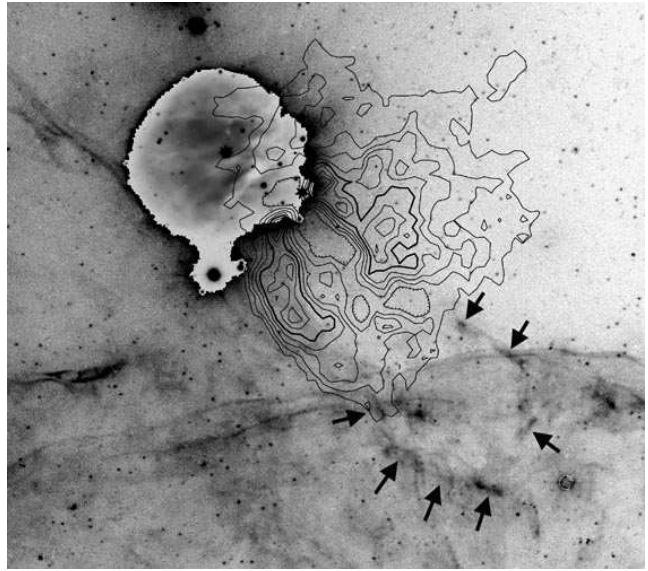}
  \caption{The  CO chimney -- H$\alpha$ cavity south-west of Sh2-219. The 
  contours of velocity-integrated CO emission are superimposed on the H$\alpha$ image of Sh2-219. 
 The first contour is $5.8\K\kms$ and the contour interval is $2.9\K\kms$. To 
 highlight the H$\alpha$ filaments, most of the low-brightness stars have been subtracted 
 from the original frame, and the filaments were digitally enhanced. The 
  arrows point to filaments forming the walls of the H$\alpha$ cavity closing the 
  molecular chimney.}
\end{figure}

\subsection{General discussion}

Sh2-219 is an almost perfectly spherical \Hp\ region 
around a central exciting star. Hence this region has probably evolved during 
most of its lifetime in a rather spherically-symmetric environment. The 
ionized gas is surrounded by an \HI\ photodissociation region (PDR), except in 
the south-west. In Sh2-219 the density 
of the ionized gas is in the range 55--170~cm$^{-3}$ (Deharveng et al.~\cite{deh03a}), 
and, as in all \Hp\ regions, the electron temperature is of the order of $10^4$~K. 
The surrounding atomic gas is of lower density, 9~cm$^{-3}$, and of lower 
temperature, 100--500~K (Hosokawa \& Inutsuka~\cite{hos05}). Thus the pressure 
in the ionized zone is higher than in the neutral surrounding zone; 
the \Hp\ region is presently still in expansion, the ionization front 
eating away the atomic ring from the inside. According to 
Roger \& Leahy~(\cite{rog93}) the \HI\ ring 
is not expanding; thus it is probably in pressure equilibrium with a lower 
density and partially ionized external medium.

Using the very simple model of \Hp\ region evolution by Dyson \& Williams~(\cite{dys97}), 
we calculate that if Sh2-219 formed and evolved in a uniform medium of density 
$10^2\cmmt$, its present radius of 2.2~pc indicates a dynamical age of 
$1.5\times 10^5$~yr. This age estimate is {\it very uncertain}, because the 
assumption of evolution in a strictly uniform medium is unrealistic. 
On the one hand, the exciting star was probably formed 
in a much denser molecular core; during that early period the region would have 
evolved more slowly than suggested by the simple model.  But on the other hand, we 
know that the \Hp\ region is expanding into a lower density medium and is 
therefore presently evolving more rapidly.

Sh2-219 is surrounded by a large atomic photodissociated region. 
Roger \& Dewdney (\cite{rog92}) and Diaz-Miller et al.~(\cite{dia98}) have presented 
models of the time-dependent evolution of photoionized and photodissociated regions 
formed by stars of various effective temperature embedded in molecular material of 
uniform density. These models do not include hydrodynamics. They show  
that the photodissociated region is relatively more extended with respect to the 
ionized region when the exciting star has an effective temperature 
below 30\,000~K (i.e. for B0--B4 stars), and that it can persist 
a long time in a low-density medium. This applies well to Sh2-219. However these 
models are static (they assume that a dissociation equilibrium has been reached); the H$_2$ 
reformation time is very long in a low density medium, much longer than the dynamical  
timescale. Thus, as Diaz-Miller et al. (\cite{dia98}) have strongly emphasized, these 
models are not adapted to regions evolving in a low density medium, 
as is the case of Sh2-219. We therefore cannot apply these models to Sh2-219 to 
estimate its age from the relative sizes and masses of the ionized and atomic regions.

Hosokawa \& Inutsuka (\cite{hos05}) present new models where   
the dynamical evolution of an \Hp\ region and of its PDR is analysed by calculating the 
radiation transfer and the thermal and chemical processes {\it with a time-dependent 
hydrodynamics code}. They illustrate their models with the case of an 
O6 star evolving in a 10$^3$~cm$^{-3}$ uniform medium, which applies well  
to Sh2-104 (Deharveng et al.~\cite{deh03b}) 
but not to Sh2-219. Hosokawa et al.~(in preparation) have improved these  
models by considering evolution in a non-uniform medium, and by taking into account 
photodissociation by the far-UV Galactic background radiation. Hosokawa et al. 
have considered the case of Sh2-219. Their model assumes that its exciting star formed 
in a cloud composed of a small high-density core (0.07~pc, 10$^5$~cm$^{-3}$) surrounded 
by a region of radially decreasing ($\propto r^{-1.5}$) density. 
The model accounts for the size of the ionized region and of the photodissociation region, 
and for the densities in the ionized and atomic material. The size of Sh2-219 implies 
an age $\sim 10^5$ yr. The CO molecule is completely photodissociated in the PDR; 
this explains why no CO emission is detected around Sh2-219 (except in the south-west).\\

We can imagine two widely different scenarios to account for the overall morphology 
of this region and its star formation history.
\begin{itemize}
\item Scenario 1: The exciting star of Sh2-219 formed inside the near-IR cluster, and 
consequently has the same age. It was subsequently ejected from the cluster following 
dynamical 
interactions (for example this may result from a three body encounter implying a binary 
system and a third star; Kroupa~\cite{kro00}, \cite{kro01}). If the age of the cluster 
is $10^{5}$~yr ($10^{6}$~yr), the present location of the exciting star 
with respect to the cluster implies an ejection velocity 
$\geq$~20~km~s$^{-1}$ ($\geq$~2~km~s$^{-1}$). Several points argue against 
this scenario: i) why is it the most massive star of the cluster which has been 
ejected? ii) why is the present radius of the Strömgren sphere 
(which depends on several independent parameters) equal to the 
distance covered by the star since its ejection? iii) the star is presently surrounded 
by a spherical \Hp\ region and by an almost spherical atomic PDR; how is this 
spherical symmetry compatible with an ejection (a highly non-isotropic process)? iv) the 
observed density of the ionized gas is higher than the observed density of the 
surrounding atomic material; if the star has been ejected from its dense parental  
molecular cloud into a low-density medium, how can we account for this? 
Also, it is difficult to eject a massive object from a cluster. It requires, first, 
mass segregation, which leads to a high density of massive objects in the very centre of 
the cluster. Ejection of massive objects generally happens several megayears 
after cluster formation, as shown by Kroupa~(\cite{kro00}, his fig.~6).

\item Scenario 2: The exciting star of Sh2-219 formed first, in a molecular condensation  
close to --~but distinct from~-- the cloud presently observed. It formed an \Hp\ region 
and photodissociated most of the surrounding molecular material. During its expansion 
the \Hp\ region interacted with a nearby pre-existing molecular cloud, 
forming a layer of  shocked compressed material at the surface of the cloud; this 
material forms the two molecular condensations presently observed 
at the southwestern border of the Sh2-219 \Hp\ region. Hosowaka et al.'s model predicts 
that, if the average density of the molecular cloud is $8\times 10^3$~cm$^{-3}$, 
the time scale of the layer fragmentation is comparable with the estimated age of Sh2-219. 
Therefore triggered star formation seems possible, the observed  
cluster being a second-generation cluster whose formation was triggered by the expansion 
of the Sh2-219 \Hp\ region. It contains one massive star exciting the UC \Hp\ region. 
A Herbig Be star -- the H$\alpha$ emission-line star, or another 
still-hidden object -- subsequently 
modified the structure of the adjacent molecular cloud, its strong wind carving a 
chimney through it. Here again several points argue against this scenario: 
i) we are unable to prove that the cluster is younger than the exciting star of Sh2-219. 
Indeed, the absence of maser emission and the small fraction of stars presenting 
a near-IR excess seem to indicate that the cluster is possibly older than 10$^5$~years; 
ii) here again the spherical symmetry of the ionized region is difficult to reconcile 
with the interaction, in one direction, with a nearby molecular cloud.
\end{itemize}

What observations would allow us to choose between these two scenarios? A measurement of the 
proper motion of the exciting star of Sh2-219, to confirm or deny the ejection, would 
be most helpful; but this measurement is not presently possible, this region lying  
too far away. Deeper near- and mid-IR frames of the cluster would be helpful, allowing an 
estimate of its age via the fraction of objects with near-IR excesses, and via the presence 
or absence of deeply embedded (class 0 or class I) objects. Spitzer IRAC 
observations of this region would probably give this information. High resolution 
observations of the two CO clumps adjacent to the \Hp\ region could confirm and give 
details about the interaction between the expanding \Hp\ region and the molecular cloud.

\section{Conclusions}
  
The molecular CO observations presented here have shown the presence of a molecular cloud 
adjacent to the \Hp\ region  and filling a `gap' in the
ring of atomic hydrogen surrounding the ionized gas. This cloud has a mass of about
2000~$M_{\odot}$. A most remarkable aspect of its morphology is a chimney, $\sim$5~pc
long, crossing the entire cloud. The Sh2-219 \Hp\ region lies at one
extremity of this chimney. The other extremity is closed off by  
filamentary H$\alpha$ walls. We suggest that a bright H$\alpha$ 
emission star, located near the ionization front, may be at
the origin of the strong wind or flow which has carved out this chimney.

The Sh2-219 \Hp\ region is in interaction with this adjacent molecular cloud. Two molecular 
condensations are observed at the interface between the \Hp\ region and the cloud. 
A near-IR cluster is observed in the direction of this interaction region. It contains 
highly reddened stars, a luminous H$\alpha$ emission star which is also a 
near-IR excess object, and at least one early B star exciting an unresolved 
ultracompact \Hp\ region. However no CO outflow and no maser emission are observed in the
direction of the cluster, suggesting that it is not very young.

Sh2-219 is one of the \Hp\ regions we proposed as candidates for the collect and 
collapse process of star formation (Deharveng et al.~\cite{deh05}). We 
now know that star formation does not result 
from this process, as the observed molecular cloud probably pre-existed, and was  
not formed of material collected during the expansion of the \Hp\ region. No completely  
satisfying explanation has been found to explain massive-star formation in this
region. 

\begin{acknowledgements}

We gratefully thank Takashi Hosokawa for constructing a completely new 
model to account for the observations of Sh2-219. We also thank D.~Gravallon and 
S.~Ilovaisky for obtaining H$\alpha$ frames of Sh2-219 in service mode at the 
Observatoire de Haute-Provence. The suggestions of our referee were helpful.   
The IRAM \mbox{30-m} Observatory staff are thanked for their support during 
the observations. This work is partially based on observations made with the  
Telescopio Nazionale Galileo (TNG) operated on the island of La Palma by 
the Fondazione Galileo Galilei 
of the Istituto Nazionale di Astrofisica at the Spanish Observatorio del 
Roque de los Muchachos of the Instituto de Astrofisica de Canarias. Observations 
have been obtained with the VLA at NRAO; the National Radio 
Astronomy Observatory is a facility of the National Science Foundation  operated 
under cooperative agreement by Associated Universities, Inc. This work has made 
use of the Simbad astronomical database operated at CDS, Strasbourg.

\end{acknowledgements}


{}
\end{document}